\documentclass{aastex631}
\usepackage{geometry}               
\usepackage{graphicx}
\usepackage{amssymb}
\usepackage{epstopdf}
\usepackage{fullpage}
\usepackage{natbib}
\usepackage{subfigure}
\usepackage{epsfig}
\usepackage{multirow}
\usepackage[figuresright]{rotating}
\usepackage{amsmath}

\def\aj{AJ}                   
\def\apj{ApJ}                 
\def\apjl{ApJL}                
\def\aap{A\&A}                

\def\icarus{Icarus}           

\def\mnras{MNRAS}             
\def\pasp{PASP}               
\def\ssr{Space~Sci.~Rev.}     
\def\nat{Nature}              

\def\jgr{J.~Geophys.~Res.}    

\def\pnas{P. Natl. A. Sci.}   
\def\pasp{Publ. Astron. Soc. Pac.}   

\def\ija{Int. J. Astrobiology}        
\def\jm{J. Meteorol.}                   
\def\jas{J. Atmos. Sci.}                

\begin{document}
\title{Increased Surface Temperatures of Habitable White Dwarf Worlds\\ 
Relative to Main-Sequence Exoplanets\\} 


\author{Aomawa L. Shields}\thanks{corresponding author}
\affil{Department of Physics and Astronomy \\
University of California, Irvine \\
4129 Frederick Reines Hall \\
Irvine, CA\\ 
92697-4575 USA}

\author{Eric T. Wolf}
\affil{Laboratory for Atmospheric and Space Physics\\
University of Colorado, Boulder\\
Boulder, CO 80303 USA}

\author{Eric Agol}
\affiliation{Department of Physics and Astronomy\\
University of Washington\\
Seattle, WA\\
98105-6698 USA}

\author{Pier-Emmanuel Tremblay}
\affiliation{Department of Physics\\
University of Warwick\\
Coventry CV4 7AL\\
UK}



\begin{abstract}
\noindent Discoveries of giant planet candidates orbiting white dwarf stars and the demonstrated capabilities of the James Webb Space Telescope bring the possibility of detecting rocky planets in the habitable zones of white dwarfs into pertinent focus. We present simulations of an aqua planet with an Earth-like atmospheric composition and incident stellar insolation orbiting in the habitable zone of two different types of stars\textemdash a 5000 K white dwarf and main-sequence K-dwarf star Kepler-62 with a similar effective temperature\textemdash and identify the mechanisms responsible for the two differing planetary climates. The synchronously-rotating white dwarf planet's global mean surface temperature is 25 K higher than that of the synchronously-rotating planet orbiting Kepler-62, due to its much faster (10-hr) rotation and orbital period. This ultra-fast rotation generates strong zonal winds and meridional flux of zonal momentum, stretching out and homogenizing the scale of atmospheric circulation, and preventing an equivalent build-up of thick, liquid water clouds on the dayside of the planet compared to the synchronous planet orbiting Kepler-62, while also transporting heat equatorward from higher latitudes. White dwarfs may therefore present amenable environments for life on planets formed within or migrated to their habitable zones, generating warmer surface environments than those of planets with main-sequence hosts to compensate for an ever shrinking incident stellar flux.
\end{abstract}
\keywords{planetary systems---radiative transfer---stars: white dwarfs---astrobiology}

\pagebreak
\section{Introduction} \label{sec:intro}
White dwarf stars are the final, electron-degenerate core phase of evolution for most stars in the Galaxy. During the dramatic red giant phase of stellar evolution, the outer shell of a lower-mass main-sequence star blows outwards, presumably engulfing any planets in or interior to the habitable zone of their progenitor, and leaving behind a small remnant stellar core. To date only giant planet candidates have been discovered around white dwarfs \citep{Luhman2011,  Gansicke2019, Vanderburg2020, Blackman2021, Mullally2024}. Potentially rocky planets have also been found at large orbital distances from the white dwarf in a system containing a pulsar, a rocky planet, and a white dwarf \citep{Thorsett1993, Sigurdsson2003}, and a system containing a white dwarf, rocky planet, and a brown dwarf \citep{Zhang2024}. Detections of circumstellar debris disk material at close and moderate orbital distances from white dwarfs \citep{Jura2003, Vanderburg2015, Vanderbosch2020, Farihi2022, Swan2024, Aungwerojwit2024} bolster the possibility that small, rocky planets may orbit in the habitable zones of these post-main sequence stars \citep{Veras2021}, even though these planets may be rare \citep{Kipping2024}, having arrived subsequent to the red giant phase (e.g., \citealp{Nordhaus2010}), either through inward migration \citep{Debes2002} or forming from the gas surrounding the white dwarf \citep{Livio2005}. 

Unlike main-sequence stars, white dwarfs continue to cool with time, resulting in a white dwarf habitable zone (WDHZ) that is extremely close to the star and inwardly migrating, while sustaining planets within its boundaries for up to 8 Gyr \citep{Agol2011b, Agol2011a}. However, long-term habitability in these systems depends on several factors. The most relevant habitability factors include the age and cooling time of the white dwarf \citep{Ostriker1968}, the incident stellar radiation at close orbital distances, which could cause orbiting planets to lose their water inventories to space in a runaway greenhouse phase at or near the inner edge of the WDHZ \citep{Barnes2013a}, the UV radiation dosage and surface environment, and requisite shielding of an orbiting planet's atmosphere (see, e.g., \citealp{Fossati2012, Kozakis2018, Gertz2019}), and the location of a planet within the WDHZ relative to the white dwarf's roche limit, within which tides will deform and possibly disrupt an orbiting planet \citep{Williams2003}.

The climate and habitability of planets are also affected by their rotation rate (see, e.g., \citealp{Edson2011, Kite2011, Showman2013, Hu2014, Wang2014a, Wang2014b, Kaspi2015, Guzewich2020, Lobo2023}), as well as the interaction between the spectral energy distribution (SED) of their host stars and the wavelength-dependent albedos of planetary surfaces \citep{Shields2013, Shields2014, Shields2018, Rushby2019, Rushby2020, Palubski2020}. Given the close orbital distances of WDHZ planets, such planets are highly likely to be synchronously-rotating, with much faster orbital and rotation periods compared to synchronously-rotating habitable-zone planets orbiting brighter, main-sequence stars, leading to major differences in atmospheric circulation and heat distribution around the planet, depending on atmospheric composition (see, e.g., \citealp{Wordsworth2010a}). The SEDs of white dwarfs, which have ceased to undergo core nuclear fusion (apart from varying degrees of residual thermonuclear burning in the hydrogen-rich envelopes of low-metallicity (Z ${<}$ 0.001) white dwarfs, see, e.g., \citealp{Chen2021}), will also differ from those of main-sequence stars with equivalent effective temperatures. Additionally, the atmospheric composition of white dwarfs can change substantially as they cool, resulting in an evolution of their spectral properties and appearance over time \citep{Bedard2020, Bedard2022}. As such, the interaction between the white dwarf SED and an orbiting planet's surface, and the resulting climatic effect on the planet, will also differ correspondingly. Studies have identified an infrared (IR) deficit in select samples of the white dwarf stellar population \citep{Bergeron1995a, Oppenheimer2001}, which could affect the interaction between the white dwarf SED and any water ice and snow on orbiting planets, as these surfaces have wavelength-dependent albedo properties, with much lower albedos in the IR \citep{Dunkle1956}. The effect of white dwarf host stars on the climate and habitability of potential orbiting Earth-sized planets has not been widely explored. Given the potential capabilities of the James Webb Space Telescope (JWST, \citealp{Gardner2006, Kalirai2018}) to characterize atmospheres of Earth-sized planets around white dwarfs \citep{Loeb2013, Kaltenegger2020, Lin2022}, quantifying the climatic impact on surface habitability of the specific stellar environment of these planets is of imminent relevance.

In this work we explore and compare the climate of a simulated aqua planet (no land) with an Earth-like atmospheric composition and incident stellar insolation (hereafter ``instellation'') orbiting in the habitable zone of a 5000 K pure-H atmosphere white dwarf with that of a planet receiving an equivalent instellation from a main-sequence star with a similar effective temperature, Kepler-62 (4859 K).\footnote{The simulated Kepler-62 planet corresponds to a hypothetical planet whose orbit would lie in between those of Kepler-62e and Kepler-62f.} While the different formation and evolutionary histories of these two planets would likely result in different interior and atmospheric compositions, their compositions are assumed to be equivalent in this study, to isolate the climatic effects of the planetary rotation and orbital period given the habitable zone location of each host star. ln Section 2 we present and explain our methods and models used to undertake this study. In Sections 3 and 4 we present and discuss the results and significance of our simulated planets' climates in the context of their habitability potential given their different stellar host environments. In Section 5 we offer concluding remarks and implications of this work for future studies of the potential climates of terrestrial planets discovered around white dwarf stars.

\section{Methods and Models} \label{sec:models}
We used the Community Earth System Model (CESM) version 1.2.1, a three-dimensional (3D) global climate model (GCM) developed to simulate past and present climate states on the Earth \citep{Gent2011}, with ExoCAM \citep{Wolf2022}, which is a modified version of the atmospheric component to CESM, the Community Atmosphere Model (CAM4), and the Los Alamos sea ice model (CICE version 4; \citeyear{Hunke2008}). ExoCAM employs a correlated-k radiative transfer code (ExoRT), and a finite-volume dynamical core. A complete description of the code and its lineage is available in Wolf \textit{et al.} (\citeyear{Wolf2022}). The ocean is treated as static, but fully mixed with depth. Simulations that include a fully dynamic ocean are too computationally expensive to permit the in-depth exploration of the forcing parameters we prioritize in this work. The horizontal resolution is 4$^\circ$$\times$5$^\circ$. We increased the horizontal resolution to 2$^\circ$$\times$2.5$^\circ$ in select sensitivity tests to confirm that Rossby waves were resolvable at the rotation rate of the white dwarf (WD) planet ($\sim$10 hr), and calculated that the climates were equivalent to those run at the lower resolution. We therefore ran our full suite of simulations at 4$^\circ$$\times$5$^\circ$ to reduce computational expense. Simulations of planets with similar rotation periods were run at this resolution in previous work \citep{Komacek2019b}. We simulated the climates of aqua planets (no land) orbiting two stars with similar effective temperatures, a modeled white dwarf (5000 K) and Kepler-62 (4859 K). We chose this effective temperature for the WD because 5000 K is near the peak of the white dwarf luminosity distribution \citep{Winget1987}, the duration within the WDHZ is at a maximum ($\sim$8 Gyr) at the 10-hr orbital period of a planet receiving Earth-like instellation from a WD at this temperature, and such a planet would be well outside of the Roche limit \citep{Agol2011b, Agol2011a}.


We created a synthetic spectrum of Kepler-62 (K62) using a surface gravity log (g)=4.59, metallicity Fe/H = $-$0.34, and effective temperature $T_{\rm eff}$=4859 K from the NASA Exoplanet Archive \footnote{\url{https://exoplanetarchive.ipac.caltech.edu/overview/Kepler-62/}}. The mass $M$ = 0.697 $M_\odot$, and radius $R$ = 0.707 $R_\odot$ from isochrone fitting yield a luminosity $L$ = 0.25 $L_\odot$ from Fulton and Petigura (\citeyear{Fulton2018}). The synthetic spectrum for the WD was obtained from a grid of synthetic spectra and cooling models for white dwarfs with pure hydrogen atmospheres\footnote{\url{https://warwick.ac.uk/fac/sci/physics/research/astro/people/tremblay/modelgrids/}} based on previous calculations \citep{Bergeron1995b, Kowalski2006, Tremblay2011}. The WD spectrum assumed a surface gravity $\log(g)$=8.0 and $T_{\rm eff}$=5000 K. This corresponds to an age of 5.96 Gyr, mass $M = 0.580$ $M_{\odot}$ and radius $R$=0.012 $R_{\odot}$, yielding a luminosity $L = 8.87\times10^{-5}$ $L_{\odot}$, based on the evolutionary cooling sequences provided by \citet{Bedard2020}. The synthetic spectra of both host stars are provided in Figure 1.

\begin{figure}[!htb]
\begin{center}
\includegraphics[scale=0.50]{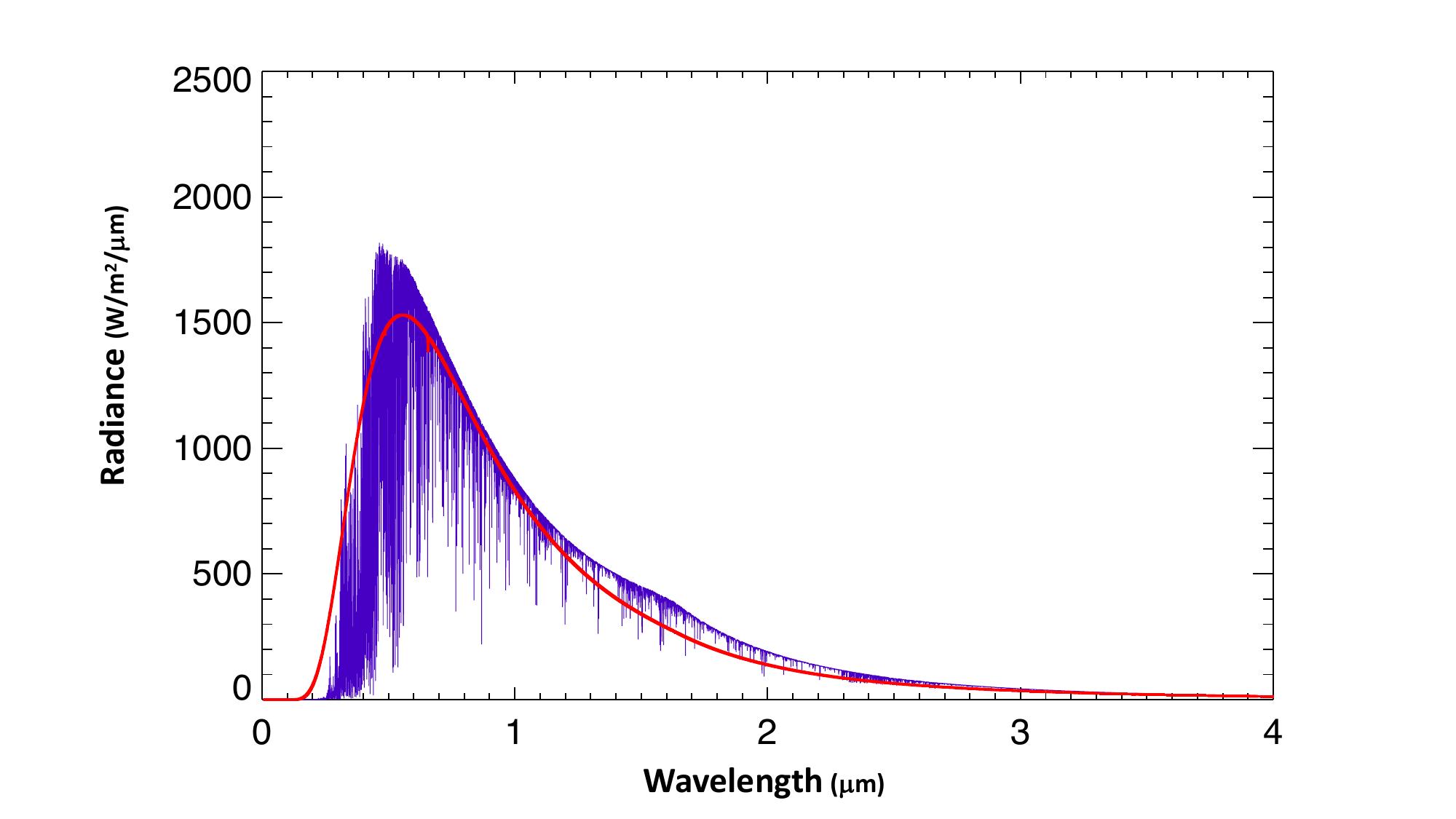}
\caption{The spectral energy distribution of a modeled white dwarf with an effective temperature of 5000 K (red) and a synthetic spectrum of Kepler-62 (4859 K, purple). Both SEDs have been normalized to an Earth-like instellation ($\sim$1360 W/m$^2$).} 
\label{Figure 1.}
\end{center}
\end{figure}


We assumed circular orbits for our orbiting planets, Earth's radius, mass, atmospheric composition, and surface pressure. The planets were assumed to be synchronously rotating, as defined as tidally-locked in a 1:1 spin-orbit resonance. Each planet was put at the distance from its host star where it would receive an Earth-like equivalent amount of instellation ($\sim$1360 W/m$^2$), given the host star's intrinsic stellar luminosity. For the simulated planets around K62 and the WD, these distances corresponded to 0.50 AU and 0.01 AU, respectively. Rotation periods were then calculated for each planet based on Kepler's third law. The majority of the simulations were carried out assuming synchronous rotation periods of 155 days and 0.44 days, assuming the K62 and WD host stars, respectively. Simulations were run until thermal equilibrium was reached, where global mean surface temperatures were stable within 1$^\circ$ over the last 20 model years.  Atmospheric water vapor was allowed to vary during each simulation according to evaporation and precipitation processes on the surface and in the atmosphere. 

Given that previous work has shown that observed planet Kepler-62f, with a 267-day orbital period, could have reached tidal circularization within the 7 Gyr-age of the system \citep{Shields2016a}, and other work also shows that it is possible with current models for our hypothetical K62 planet to orbit within the tidal locking radius \citep{Barnes2017}, it is reasonable to assume synchronous rotation for the planet in this study. However, as rotational state can vary across a range of rotation periods (see, e.g. \citealp{Yang2014}), and processes such as atmospheric tides could counteract gravitational inducements toward synchronous rotation \citep{Leconte2015}, we also simulated the potential climate of the K62 planet in a non-synchronous case, by prescribing a 10-hr rotation period as with the WD planet, while leaving its orbital period at 155 days, to allow for an additional comparison of the climates of these two planets in an equivalent rotation period scenario.


The sea-ice albedo parameterization used with ExoCAM divides the surface albedo into two bands (as done in Shields \emph{et al.} (\citeyear{Shields2013}, \citeyear{Shields2014}, \citeyear{Shields2016a})\textemdash visible ($\lambda \leqslant$ 0.751880 $\mu$m) and near-IR ($\lambda >$ 0.751880 $\mu$m)\textemdash because it is easier to control than the multiple-scattering scheme in other code versions. The default near-IR and visible band albedos are 0.3 and 0.67 for bare water ice, and 0.68 and 0.8 for dry snow, respectively, assuming the Sun as the host star. For our simulations of the climates of planets orbiting K-dwarf star K62 and the WD, we calculated the two-band albedos weighted by the specific spectrum of each host star, and used those values (Table 1) as input. 

\linespread{1.0}
\begin{table}[!htp] 
\caption{Two-band albedos employed for different temperature regimes (given in degrees Celsius) reached in the GCM, weighted by the spectrum for K-dwarf star K62 (4859 K) and a modeled WD (5000 K). $E$ and $P$ denote water evaporation and precipitation, respectively. Where $E-P<0$, albedos for dry snow are employed.} 
\vspace{2 mm}
\centering \begin{tabular}{c c c c c} 
\hline\hline 
Host star & $0^\circ>T>-23^\circ$ & $-23^\circ>T>-40^\circ$ & $T<-40^\circ$ & $E-P<0$ \\  [0.5ex] 
\hline
Band & NIR/VIS & NIR/VIS & NIR/VIS & NIR/VIS\\
K62 & 0.20/0.69 & 0.24/0.81 & 0.89/0.95 & 0.55/0.98\\ 
WD & 0.21/0.70 & 0.26/0.82 & 0.90/0.95 & 0.59/0.98\\[1ex]
\hline 
\end{tabular} 
\label{table:nonlin} 
\end{table}

Modifications were made to the ice thermal code in CICE4, based on the original model written by Bitz \emph{et al.} (\citeyear{Bitz2001}), to incorporate the bare sea ice albedo change due to the crystallization of hydrohalite at low temperatures, and the subsequent formation of a hydrohalite crust (see, e.g., \citealp{Carns2015, Light2016, Carns2016}), as done in previous simulations of exoplanet climates \citep{Shields2018}. In the model, sea ice is allowed to form as surface temperatures reach the freezing point of liquid water.  Areas of net water precipitation were assigned two-band albedos for salt-free snow with 100-$\mu$m sized grains. For temperatures between freezing and the temperature where hydrohalite begins to precipitate in sea ice (T$<$$-$23$^\circ$C), we used two-band albedos for salt-free, ``warm" bare ice; below $-$23$^\circ$C, we used two-band albedos for cold bare ice with precipitated hydrohalite; and below $-$40$^\circ$C, we used two-band albedos for a fully-formed hydrohalite crust. Two-band albedos weighted by the spectrum of each host star and used for each temperature regime are given in Table 1.



For temperature regimes where a hydrohalite crust was expected to form ($T<-40^\circ$C), we altered the emissivity from a value of 1 for both salt-free water ice and snow, as well as cold bare ice with precipitated hydrohalite (99\% H$_2$O and 1\% NaCl), to 0.752\textemdash a weighted average between that of salt-free H$_2$O snow and fine-grained halite (60\% in the thermal IR, \citealp{Lane1998})\textemdash to approximate with greater accuracy the emissivity of a hydrohalite crust (62\% NaCl and 38\% H$_2$O by weight). We found this change in emissivity to be relevant in our simulations given an Earth-like atmospheric composition, as temperatures reached below the hydrohalite crust formation threshold in the polar regions and wider on these planets, resulting in changes in their global mean surface temperatures of as much as 2 degrees K in simulations with the temperature-dependent emissivity parameterization applied.

\section{Results} \label{sec:res}
Figure 2 shows surface temperature, top-of-atmosphere albedo, surface albedo, and snow depth across synchronously-rotating planets with an Earth-like composition receiving Earth-like instellation from each star in our study. The planet orbiting K62, with its 155-day rotation and orbital period, shows a characteristic, oval-shaped temperature pattern (Fig. 2A), with the hottest point occurring at the substellar point on the planet's dayside, cooler temperatures occupying successive annuli outward from this point, and a cold nightside. In contrast, the synchronous WD planet, with its 10-hr rotation and orbital period, exhibits a distinctive pattern, with extended and stretched out scales of circulation across the planet and mid-latitude jets, with the hottest surface temperatures located in these regions, similar to what is seen in simulations of other short-period planets (see, e.g., \citealp{Haqq-Misra2018, Komacek2019a, Zhan2024}). 

\begin{figure}[!htb]
\begin{center}
\hspace{0.05mm}
\includegraphics[scale=0.40]{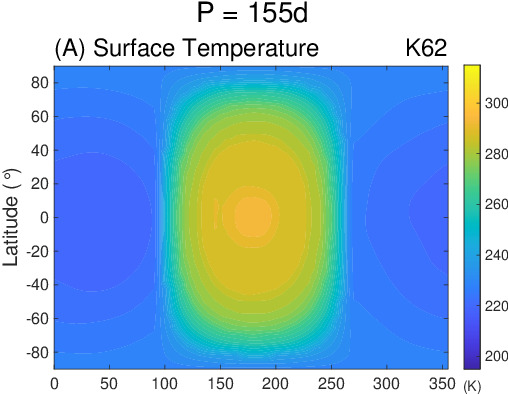}
\includegraphics[scale=0.40]{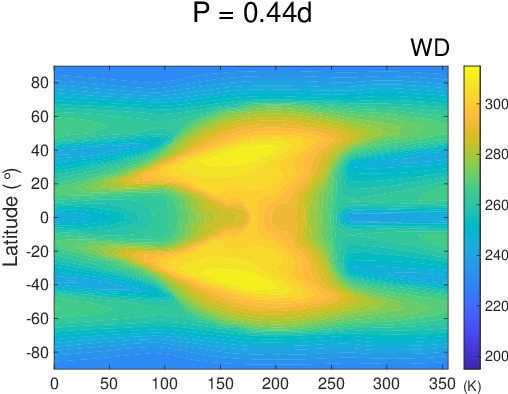}\\
\includegraphics[scale=0.40]{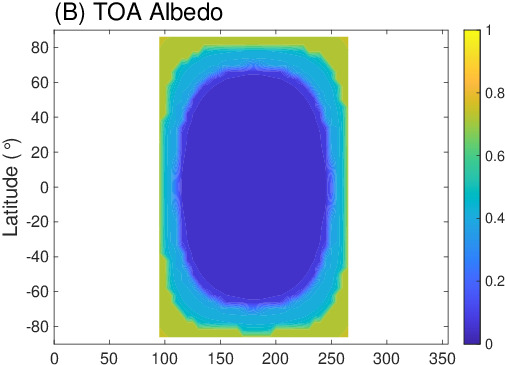}
\includegraphics[scale=0.40]{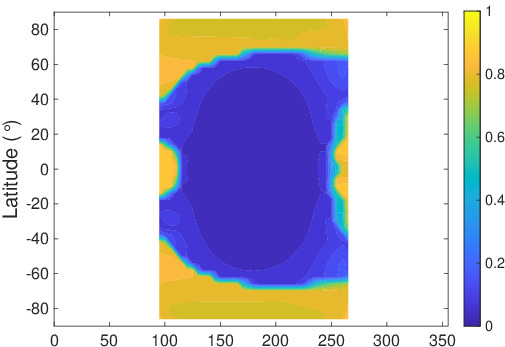}\\
\includegraphics[scale=0.40]{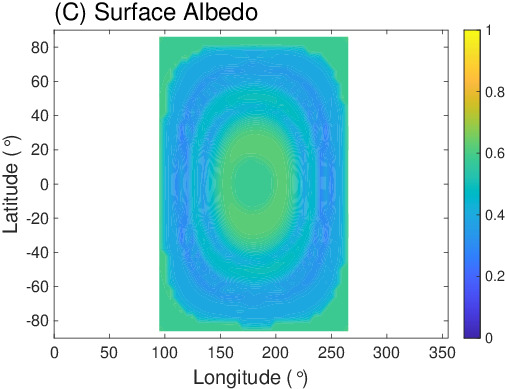}
\includegraphics[scale=0.40]{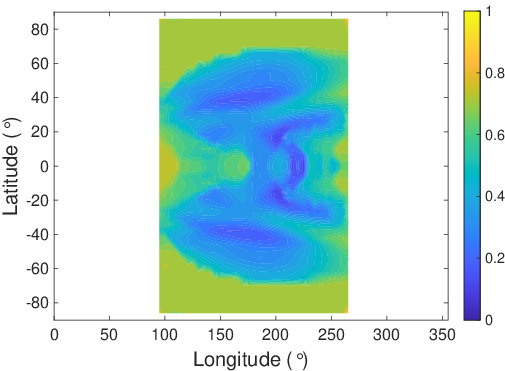}\\
\includegraphics[scale=0.40]{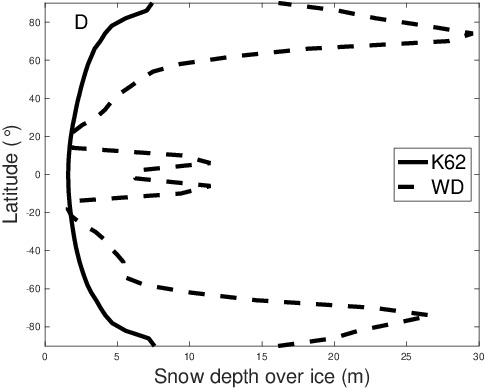}\\
\caption{Climate comparisons for synchronously-rotating planets with Earth-like atmospheric compositions receiving 100\% of the modern solar constant from Kepler-62 and a 5000 K synthetic white dwarf. Figure 2D is averaged over longitude. The slower-rotating K62 planet exhibits an oval-shaped temperature pattern with the hottest regions at the substellar point, high TOA albedo albedo due to large substellar cloud coverage, lower-albedo ice at the polar regions relative to the WD planet, and a cold nightside. The WD planet exhibits a lower TOA albedo and higher-albedo thick snow at the poles, while it's ultra-fast rotation generates a more homogeneous appearance between day and night sides, mid-latitude jets and a banded pattern, yielding hotter surface temperatures in these regions, with a global mean surface temperature $\sim$25 K higher higher than on the K62 planet.} 
\label{Figure 2.}
\end{center}
\end{figure}

The jets seen on the WD planet exhibit homogeneity in temperature across the day/night boundary. In particular, the WD planet has a jet band across the pole from day to nightside, and a minimum surface temperature at the dayside pole of the planet that is equivalent to its nightside minimum surface temperature. On the synchronous K62 planet, the nightside gets much colder than the pole of its dayside, with a minimum surface temperature that is 9 degrees cooler than the dayside minimum. The overall nightside hemisphere mean surface temperature is $\sim$36 K colder than that of the WD planet.  

On the slower-rotating K62 planet, large dayside cloud coverage centers around the substellar point, contributing to a high top-of-atmosphere (TOA) albedo (Fig. 2B).  The WD planet has a TOA albedo that is 18\% lower and is noticeably warmer, with a maximum surface temperature that is 18 degrees higher (314 K) than that on the K62 planet and a global mean surface temperature $\sim$25 K higher. As K62 is a slightly cooler star than the WD, it emits more flux at longer wavelengths, which water ice and snow absorb more strongly compared to visible and near-UV radiation (see, e.g., \citealp{Joshi2012, Shields2013, Shields2014}). The effects of the wavelength-dependent albedo of water ice are seen at the poles of the two planets, with the K62 planet exhibiting darker ice at the poles (Fig. 2C) compared to the poles of the WD planet, which have high-albedo thick snow in these regions (Fig. 2D). Relevant values for surface temperature, TOA albedo, and ice fraction are given in Table 2. 

The effects of the large difference in rotation period between the two planets are evident in the resulting atmospheric cloud patterns as shown in Figure 3. While the K62 planet has sharply more clouds on the dayside than the nightside (Fig. 3A), the WD planet has a fairly uniform cloud fraction day to night, with a similar jet-like pattern to the surface temperature pattern seen in Fig. 2B. The WD planet has a slightly larger dayside column-integrated cloud fraction than the synchronous K62 planet. However, a closer look at the contribution of individual cloud mass to the total cloud fraction reveals a different trend. In spite of the WD planet's higher surface temperatures, the synchronous K62 planet has 20$\%$ more dayside liquid water cloud mass (Fig. 3B), which contributes to the higher TOA albedo and stronger (greater net negative) incoming stellar, or ``shortwave" (SW) cloud forcing (SWCF) on the K62 planet (Fig. 3C and Table 2). 

\begin{figure}[!htb]
\begin{center}
\includegraphics[scale=0.40]{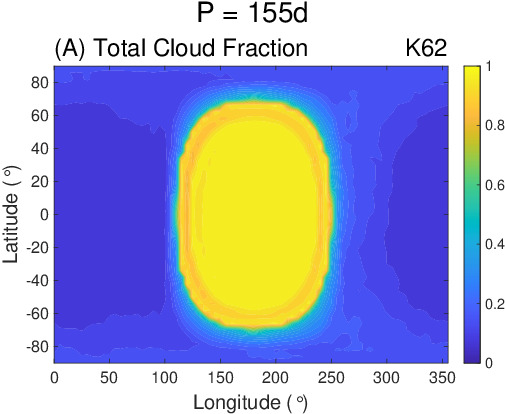}
\includegraphics[scale=0.40]{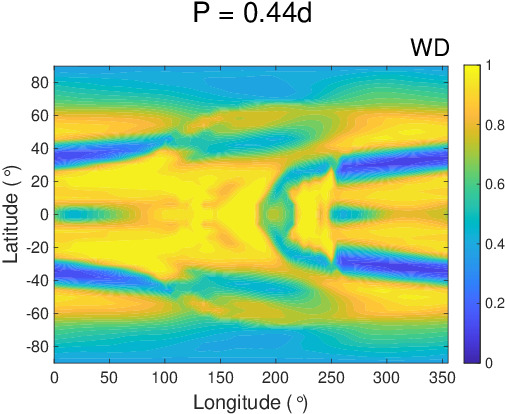}\\
\includegraphics[scale=0.40]{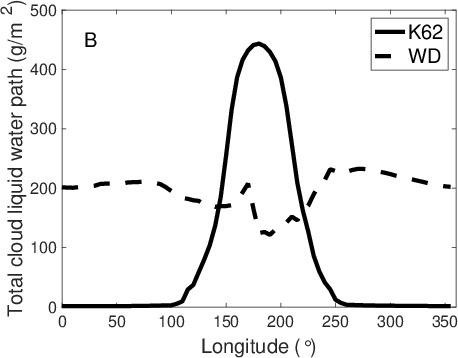}\\
\includegraphics[scale=0.40]{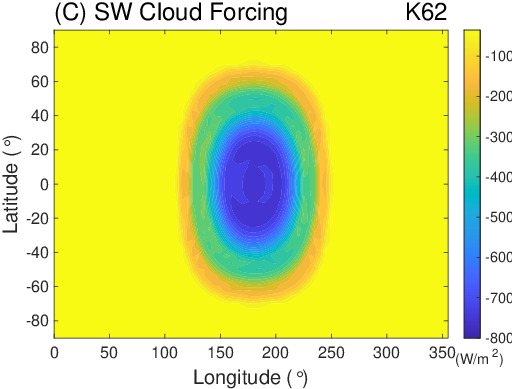}
\includegraphics[scale=0.40]{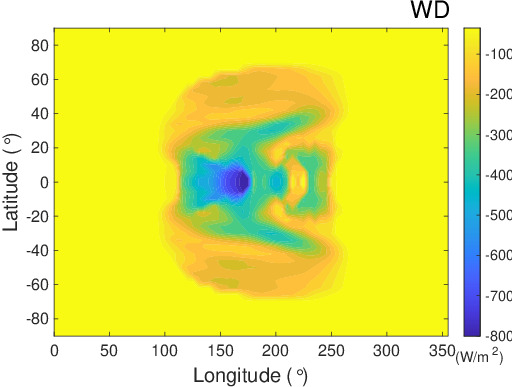}\\
\caption{Though the WD planet exhibits higher dayside surface temperatures, the K62 planet has a larger dayside cloud fraction and liquid water cloud mass (Fig. 3A and latitude-averaged Fig. 3B), contributing to a stronger (greater net negative) shortwave (SW) cloud forcing and cloud feedback (Fig. 3C), further cooling temperatures on the K62 planet.}  
\label{Figure 3.}
\end{center}
\end{figure}

The decreased cloud fraction throughout most of the atmospheric column on the dayside of the WD planet (Fig. 4A) decreases the amount of SW absorption and heating in the upper atmosphere, allowing more SW radiation to make it through the atmosphere, contributing to the increased temperatures on the surface of the WD planet (Fig. 4B). This effect is amplified further on the nightside, where the WD planet's surface is even warmer than that of the K62 planet. A comparison of the longwave (LW) cloud forcing (LWCF) on the nightside, which indicates the contribution of clouds to a planet's greenhouse effect, reveals the synchronous K62 planet to be emitting on average $\sim$18 W/m$^2$ more radiation to space on its nightside than the WD planet (Fig. 5), further reducing temperatures.

\begin{figure}[!htb]
\begin{center}
\includegraphics[scale=0.40]{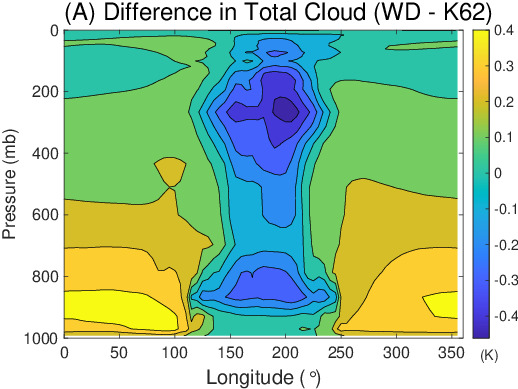}\\
\includegraphics[scale=0.40]{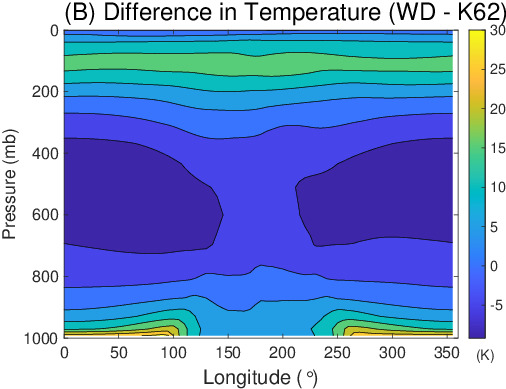}\\
\caption{The WD planet has a lower cloud fraction throughout most of the atmospheric column on the dayside compared to the K62 planet (Fig. 4A), resulting in less SW heating in these regions, allowing more radiation to make it through the atmosphere to heat the surface of the WD planet (Fig. 4B).} 
\label{Figure 4.}
\end{center}
\end{figure}

\linespread{1.0}
\begin{table}[!htp] 
\caption{Model inputs (above single solid line) and climate data (below single solid line) for simulations of Earth-sized planets orbiting Kepler-62 and a modeled 5000 K white dwarf. For synchronous rotators, the majority of the ice present, which is on the nightside, does not contribute to the surface and TOA albedo calculations, as sunlight never touches those regions. The K62 non-synchronous planet has a larger amount of ice contributing to the surface and TOA albedos than the synchronous planets' (dayside) ice, given that on the non-synchronous rotator all longitudes of the planet receive sunlight.} 
\vspace{2 mm}
\centering \begin{tabular}{c c c c} 
\hline\hline 
Host star & K62 (sync) & WD (sync) & K62 (non-sync) \\ [0.5ex] 
 Instellation (W/m$^2$) & 1361.27  & 1361.27 & 1361.27\\
 Atmospheric composition & 367 ppmv  CO$_2$ & 367 ppmv  CO$_2$ & 367 ppmv  CO$_2$\\
  &                    1.76 ppmv  CH$_4$ & 1.76 ppmv  CH$_4$ & 1.76 ppmv  CH$_4$\\
Rotation period (days)       &  155          & 0.44 & 0.44\\
Orbital period (days)		&   155	     & 0.44 & 155\\
\hline
Global mean T$_S$ (K)     & 247.8      & 273.1 & 281.9\\
Dayside mean T$_S$	& 273.2  	    & 286.7 & 282.8\\
Nightside mean T$_S$      & 223.9      & 260.2 & 281.0\\
Maximum T$_S$ 		& 296.0    & 314.2 & 314.0\\
Minimum T$_S$                & 219.5      & 230.3 & 195.8\\
TOA Albedo			& 0.4895      & 0.3999 & 0.3470\\
Surface Albedo			& 0.1675     & 0.1109 & 0.2332\\
Ice fraction			& 0.6724	    & 0.5520& 0.3053\\
Cloud fraction (Global)	& 0.4064	    & 0.7342 & 0.5812\\
Cloud fraction (dayside)    & 0.7293	    & 0.7475 & 0.5803\\
Cloud fraction (nightside)  & 0.1014      & 0.7216 & 0.5820\\
Global liquid water cloud mass (g/m$^2$) & 125.9  & 224.1 & 136.5 \\
Dayside liquid water cloud mass  & 257.8 & 213.2 & 131.2\\
Nightside liquid water cloud mass  & 1.154 & 234.5 & 141.8\\
Global ice water cloud mass  & 16.13 & 23.05 & 16.04\\
Dayside ice water cloud mass  & 32.47 & 24.55 & 15.53\\
Nightside ice water cloud mass  & 0.6916 & 21.63 & 16.54\\
SW cloud forcing (W/m$^2$)   & $-$130.7  & -90.16 & $-$61.56\\
Global LW cloud forcing            & 22.94 & 30.10 & 25.33\\
Dayside mean LW cloud forcing & 47.46 & 43.45 & 24.87\\
Nightside mean LW cloud forcing & -0.2316 & 17.49 & 25.79\\
\hline 
\end{tabular} 
\label{table:nonlin} 
\end{table}

The WD planet exhibits much stronger zonal winds and meridional flux of zonal eddy momentum compared to its synchronous counterpart orbiting K62. In contrast to the K62 planet's symmetrical pattern, strong phase tilts of the meridional flux of zonal momentum are seen above and below the equator on the WD planet (Fig. 6). These phase tilts are in the directions northeast-southwest in the northern hemisphere and northwest-southeast in the southern hemisphere. The net result of these stronger areas of phase tilt is the transport of moisture from higher latitudes towards the equator, with local minima in the regions of convergence of phase tilted meridional flux as well as along the midlatitude zonal wind jets, and local maxima along the phase tilted directions north and south of the equator, increasing surface temperatures in those regions. The substantially lower meridional flux of zonal momentum on the slow-rotating K62 planet, and minimal zonal wind strength contribute to the planet's large cloud fraction, which is relatively uniform and centered at the substellar point. 

\begin{figure}[!htb]
\begin{center}
\includegraphics[scale=0.40]{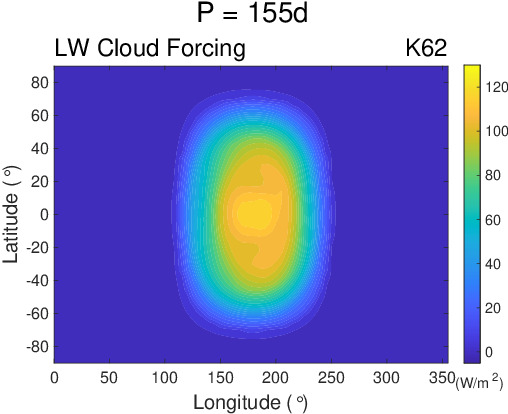}
\includegraphics[scale=0.40]{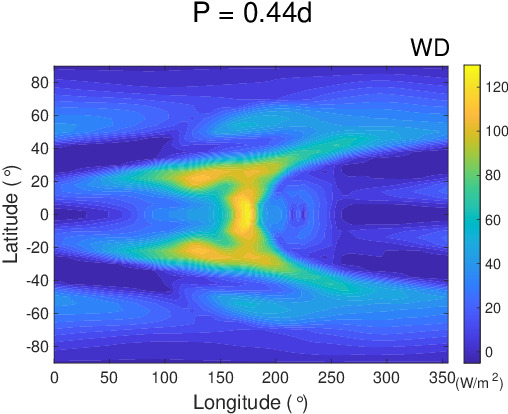}\\
\caption{A comparison of the longwave (LW) cloud forcing on the night sides of both synchronous planets reveals a much stronger contribution of clouds to the WD planet's greenhouse effect compared to the K62 planet, which emits on average $\sim$18 W/m$^2$ more radiation to space on its nightside (also see Table 2).} 
\label{Figure 5.}
\end{center}
\end{figure}

The synchronous WD and K62 planets exhibit strong and almost opposing zonal wind patterns, as shown in Figure 7. Equatorial superrotation is present throughout the atmospheric column on the K62 planet. On the fast-rotating WD planet, the maxima in zonal momentum flux westward of the substellar point inform the zonal wind pattern, which indicates westward subrotation throughout most of the atmospheric column at the equator and subtropics, up to $\sim$40$^\circ$. A small region of eastward equatorial superrotation is present from $\sim$ 500 mb down to the surface, and in the upper latitudes throughout the troposphere, similar to what is seen in other recent work, and governed largely by the hotter mid-latitudes compared to the equator, resulting in reversed low-latitude jets from eastward to westward outside of the equatorial regions \citep{Zhan2024}.

\begin{figure}[!htb]
\begin{center}
\includegraphics[scale=0.40]{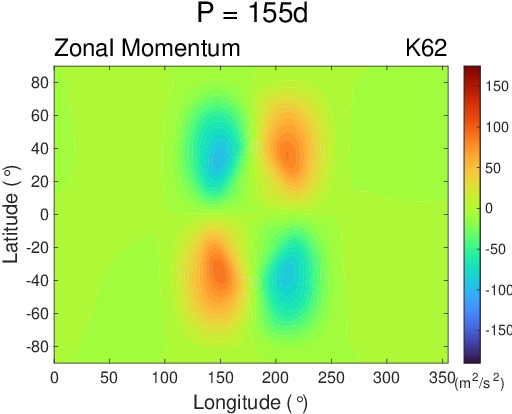}
\includegraphics[scale=0.40]{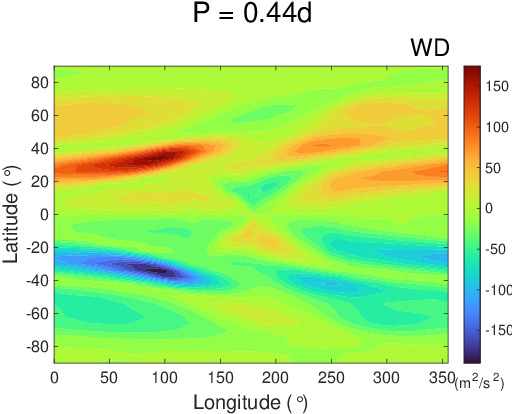}\\
\caption{The K62 planet's weaker and symmetrical meridional flux of zonal eddy momentum contributes to a large and relatively uniform cloud fraction centered at the substellar point. In contrast, strong phase tilts above and below the equator on the WD planet transport moisture from higher latitudes, with local minima in the regions of convergence and along the midlatitude zonal wind jets, and local maxima along the phase tilted directions north and south of the equator, increasing surface temperatures in those regions.} 
\label{Figure 6.}
\end{center}
\end{figure}

\begin{figure}[!htb]
\begin{center}
\includegraphics[scale=0.40]{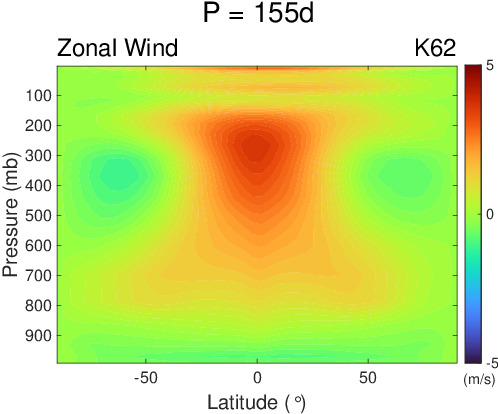}
\includegraphics[scale=0.40]{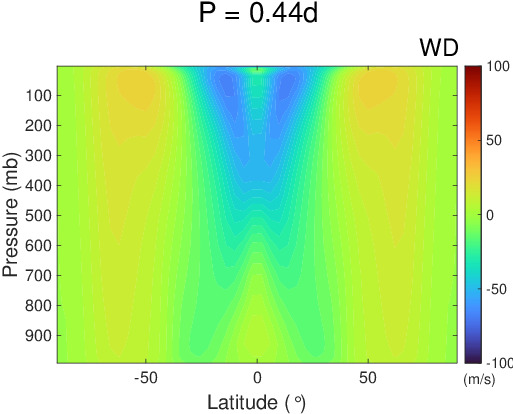}\\
\caption{Zonal wind strength is more than an order of magnitude smaller on the slow-rotating K62 planet compared to the WD planet (note different colorbar ranges). Equatorial superrotation is present throughout the atmospheric column on the K62 planet, while on the fast-rotating WD planet, westward subrotation is present throughout most of the atmospheric column at the equator and subtropics, following the maxima in zonal momentum flux. Eastward equatorial superrotation comprisies a small region of the atmosphere below $\sim$ 500 mb, and in the troposphere in the upper latitudes, given that those regions are hotter compared to the equator on the WD planet.} 
\label{Figure 7.}
\end{center}
\end{figure}

\subsection{Climate comparison with a non-synchronous K62 planet}
We also compared the resulting climates of both planets at equivalent rotation periods of 10 hrs. The WD planet is therefore synchronous, while the K62 planet is non-synchronous. The most notable differences are shown in Figure 8, with complete climate variable data listed in Table 2 alongside that of the synchronous planets.

As shown in Figure 8A, the non-synchronous planet orbiting K62, with a 10-hr rotation period and a 155-day orbital period, exhibits a surface temperature pattern quite different from either of the synchronous planets. Its pattern is longitudinally homogeneous, similar to rapidly-rotating long orbital-period planets like the Earth, with warm tropical and mid-latitude regions and cooler temperatures in the upper latitudes and at the poles. On the non-synchronous K62 planet, all longitudes of the planet receive sunlight over some portion of the planet's 10-hr day, giving rise to a global mean surface temperature that is $\sim$34 K warmer than on the synchronous K62 planet and $\sim$9 K warmer than that of the WD planet. However, a larger equator-to-pole temperature difference is evident on the non-synchronous K62 planet, with a minimum surface temperature that is $\sim$24 K lower than that on the synchronous K62 planet and $\sim$35 K lower than that of the WD planet. The non-synchronous K62 planet's diurnal instellation across all longitudes contributes to both a lower cloud fraction during the planet's day (Fig. 8B), a lower TOA albedo relative to the synchronous planets (Table 2), and a similar maximum surface temperature to the WD planet. 

Where there is ice on the non-synchronous K62 planet, at upper latitudes as shown in Figure 8C, the ice has (similar to the K62 synchronous planet) a lower albedo compared to the WD planet which, along with the much stronger LWCF on the planet's nightside (see LWCF nightside mean comparison in Table 2), aids in heat retention, contributing to the higher global mean surface temperature relative to the other two planets. The overall warmer temperatures on the WD planet and the non-synchronous K62 planet result in global mean ice fractions that are 18\% and 54\% lower than on the synchronous K62 planet, respectively. On the synchronous planets, surface ice is concentrated primarily on the nightside, therefore not contributing to the surface and TOA albedo calculations given that sunlight never touches those regions. The K62 non-synchronous planet has a larger amount of ice contributing to the surface and TOA albedos than the synchronous planets' (dayside) ice, as on the non-synchronous rotator all longitudes of the planet receive sunlight. This higher fraction of dayside ice is indicative of the much lower minimum surface temperature ($\sim$196 K) on the non-synchronous K62 planet compared to either of the two synchronous planets. The difference between maximum and minimum surface temperatures increases with decreasing rotation period for the synchronous planets, with the K62 planet exhibiting a $\sim$74 K difference, compared to an $\sim$84 K difference on the WD planet. While its global mean, nightside mean, and dayside mean surface temperatures are nearly equivalent, the non-synchronous K62 planet has the largest max-min surface temperature difference given its larger dayside ice fraction, at $\sim$118 K, as shown in Figure 9. Relevant values for surface temperature, TOA albedo, and ice fraction are given in Table 2.

\begin{figure}[!htb]
\begin{center}
\includegraphics[scale=0.29]{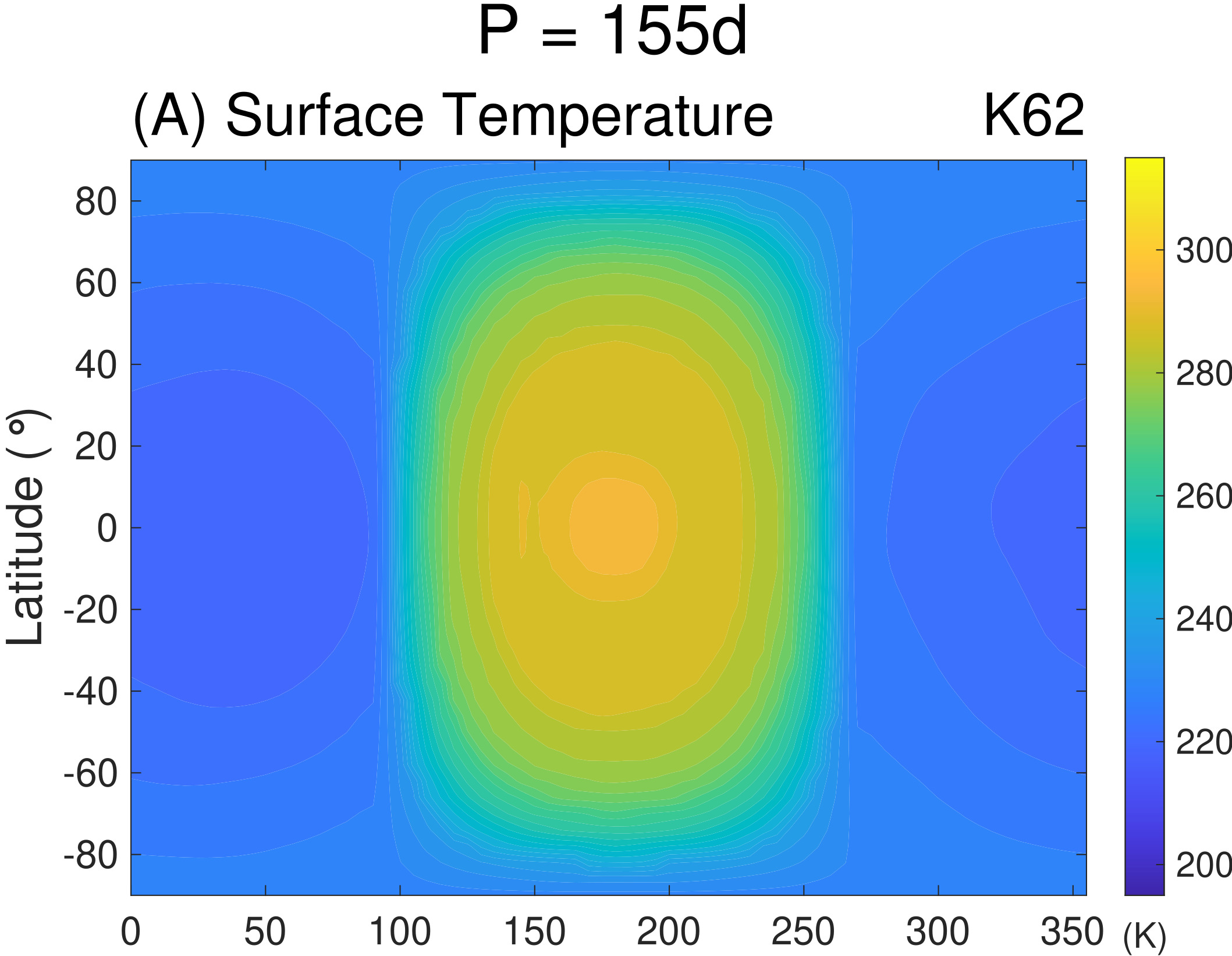}
\includegraphics[scale=0.29]{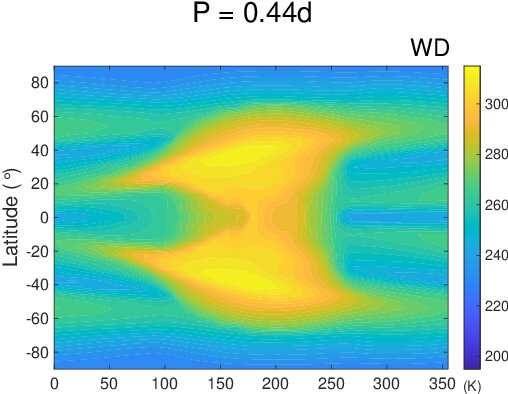}
\includegraphics[scale=0.29]{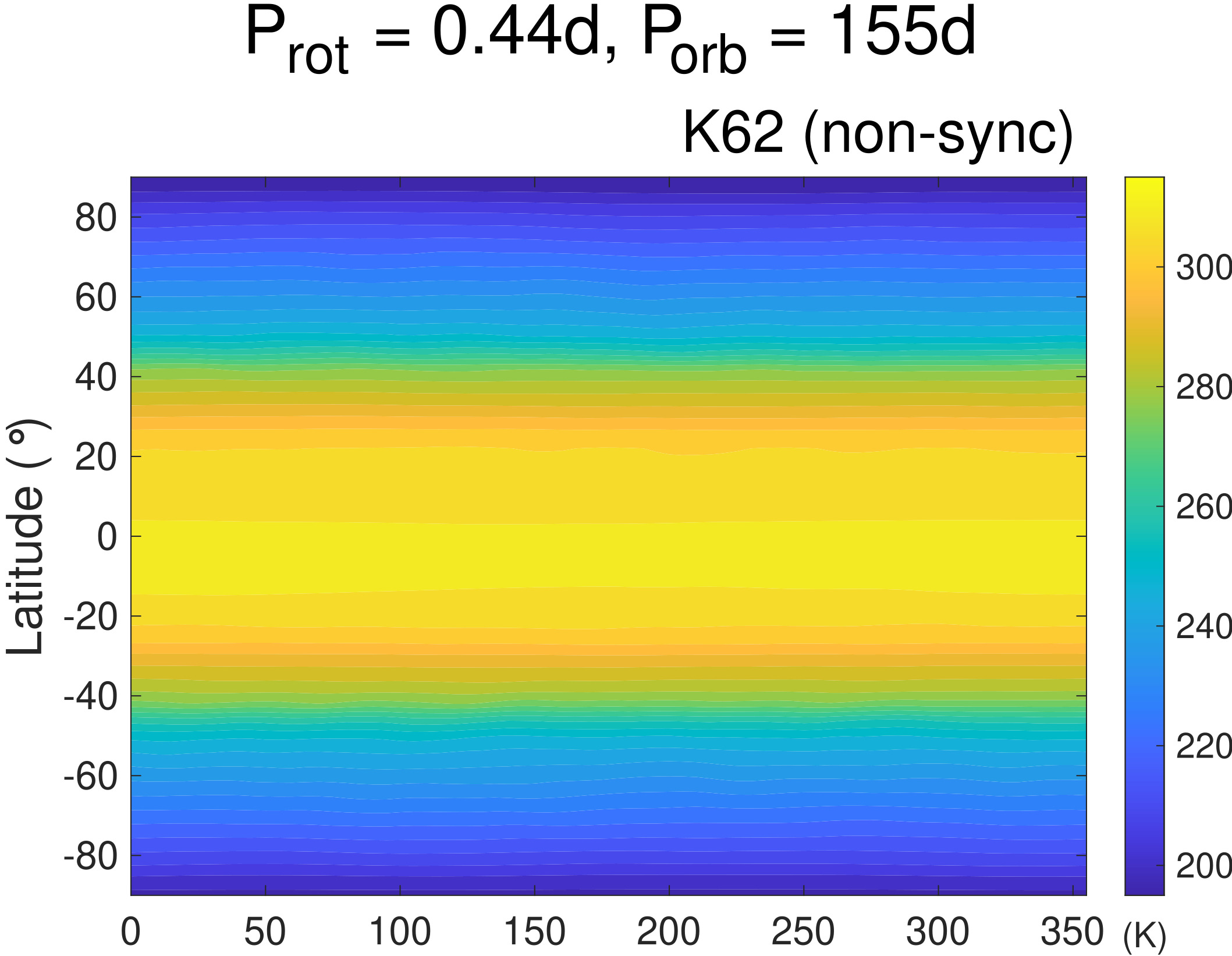}\\
\includegraphics[scale=0.29]{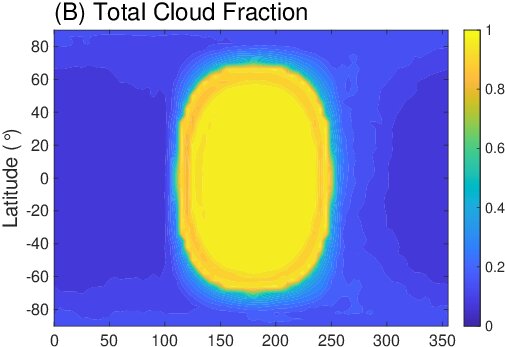}
\includegraphics[scale=0.29]{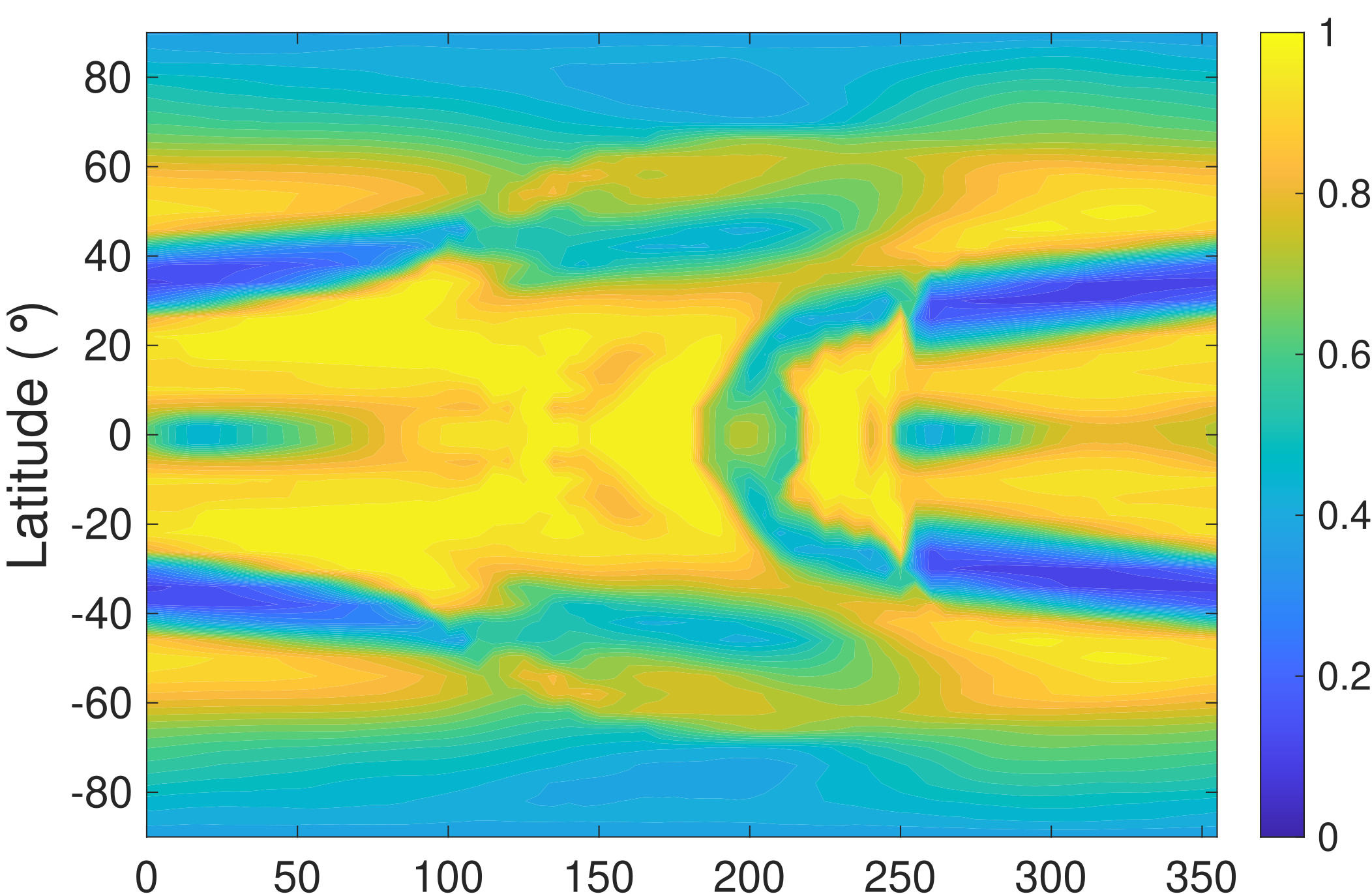}
\includegraphics[scale=0.29]{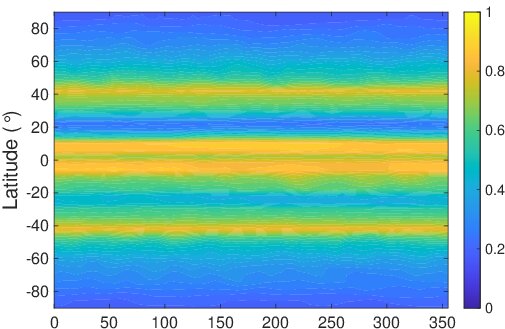}\\
\includegraphics[scale=0.29]{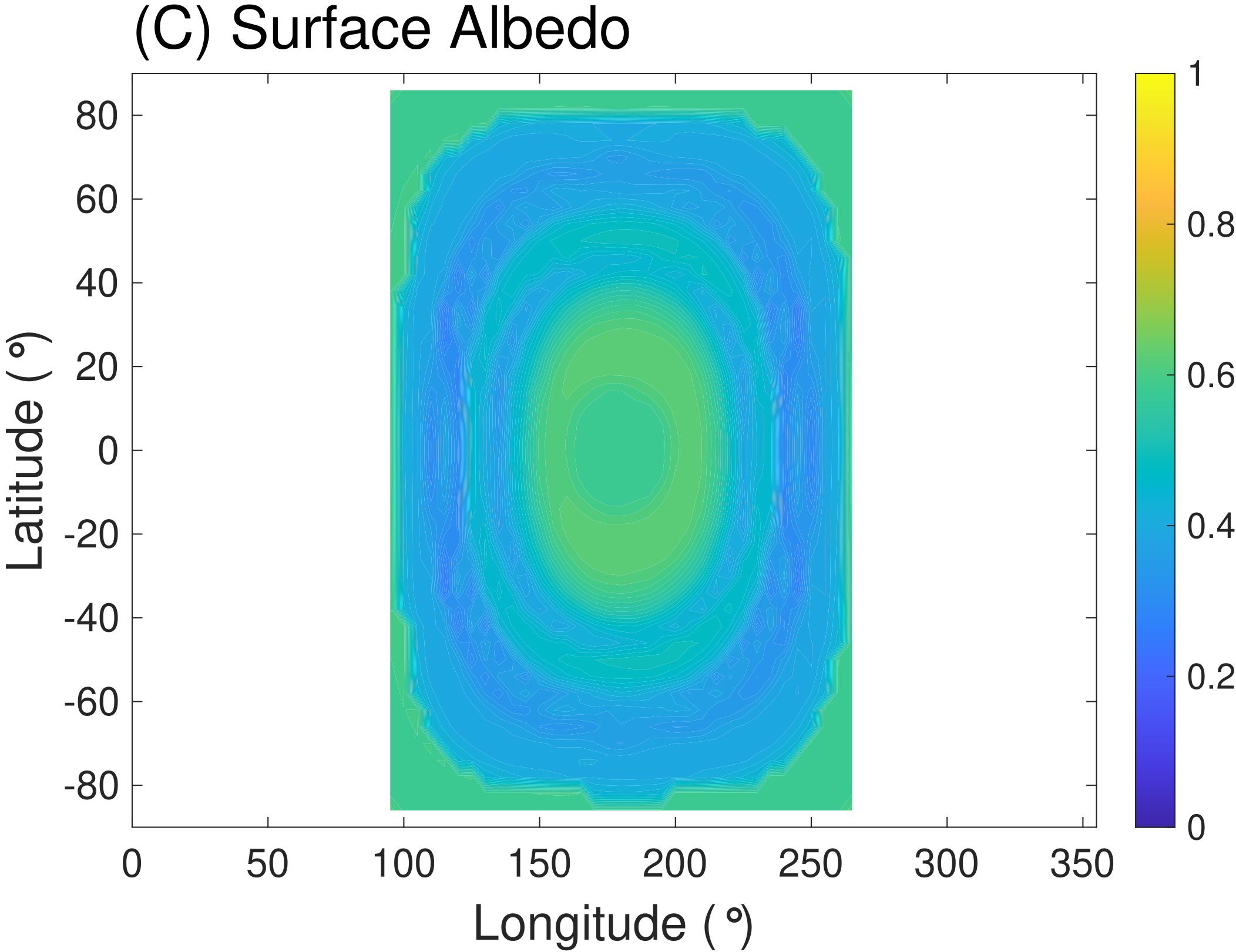}
\includegraphics[scale=0.29]{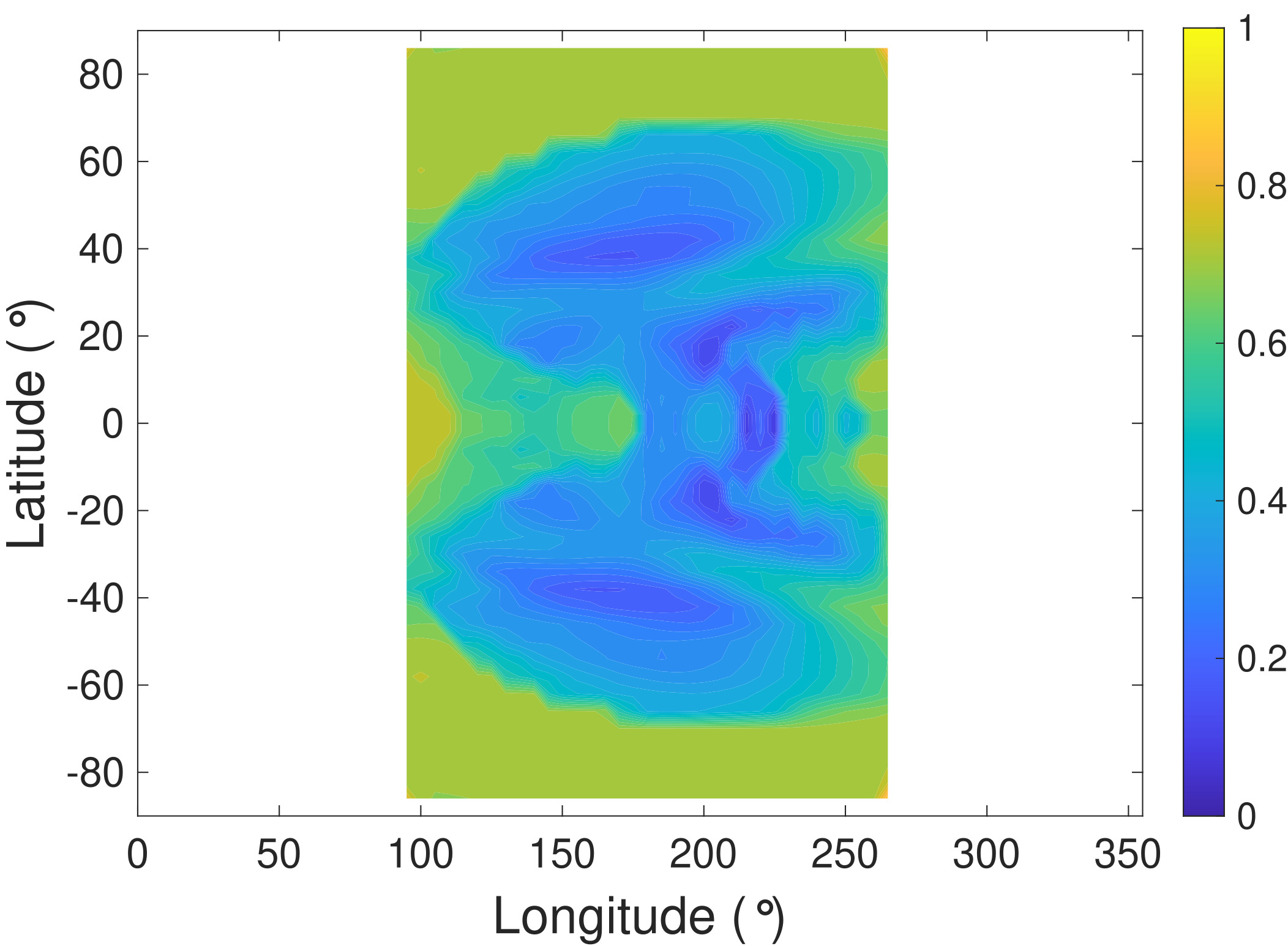}
\includegraphics[scale=0.29]{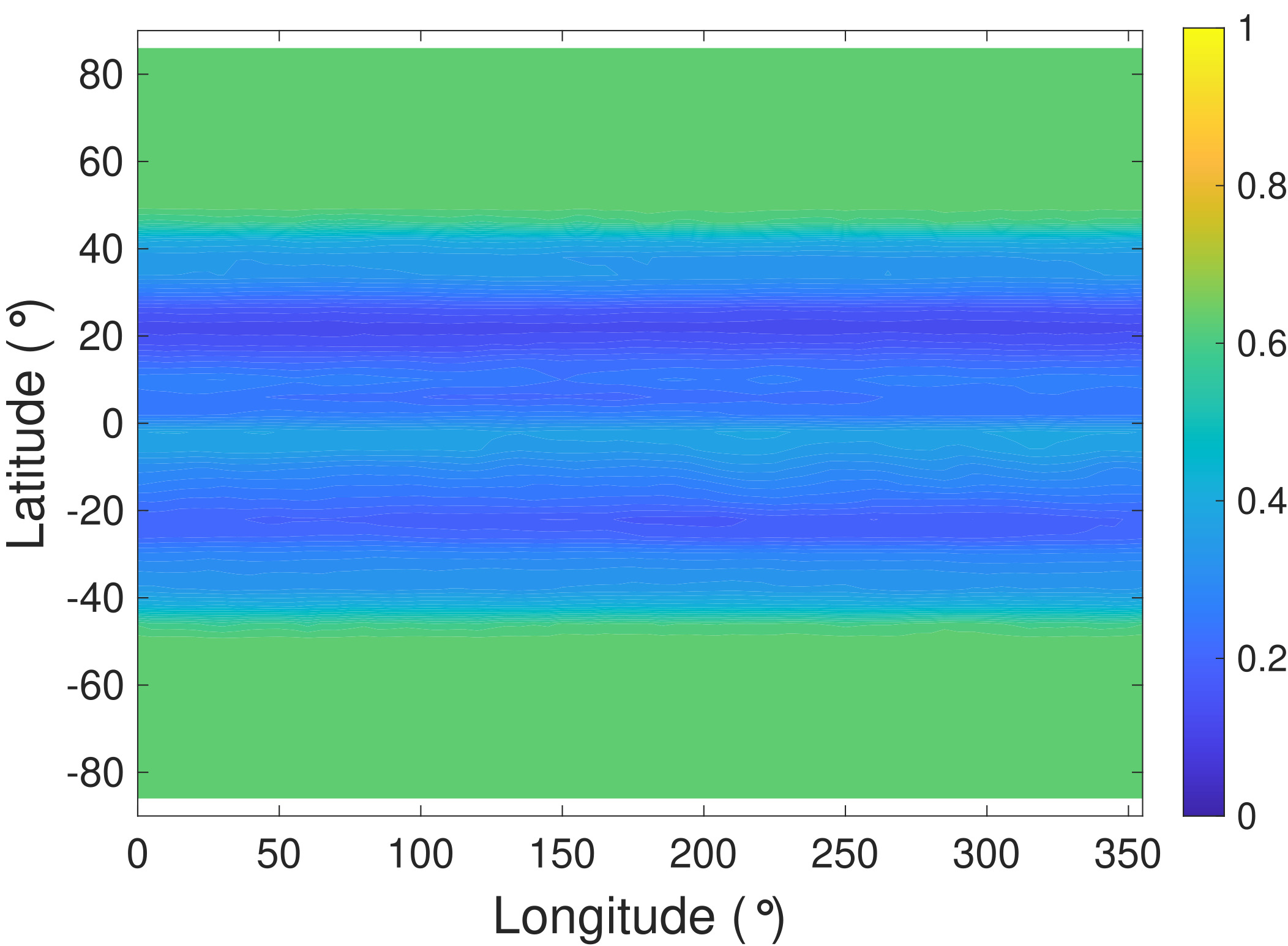}\\
\caption{Climate comparison of a non-synchronous, fast-rotating K62 planet with both synchronous planets. The non-synchronous K62 planet is warmer overall, with a surface temperature  pattern that is longitudinally homogeneous. All longitudes of the planet receive sunlight over some portion of the planet's 10-hr day, giving rise to a lower cloud fraction and a larger amount of ice contributing to the surface and TOA albedos than the synchronous planets' (dayside) ice. This higher fraction of dayside ice on the non-synchronous planet contributes to a larger equator-to-pole temperature difference compared to either of the synchronous planets, as shown in Figure 8A. Relevant values for surface temperature, TOA albedo, and ice fraction are given in Table 2.} 
 \label{Figure 8.}
\end{center}
\end{figure}

\begin{figure}[!htb]
\begin{center}
\includegraphics[scale=0.70]{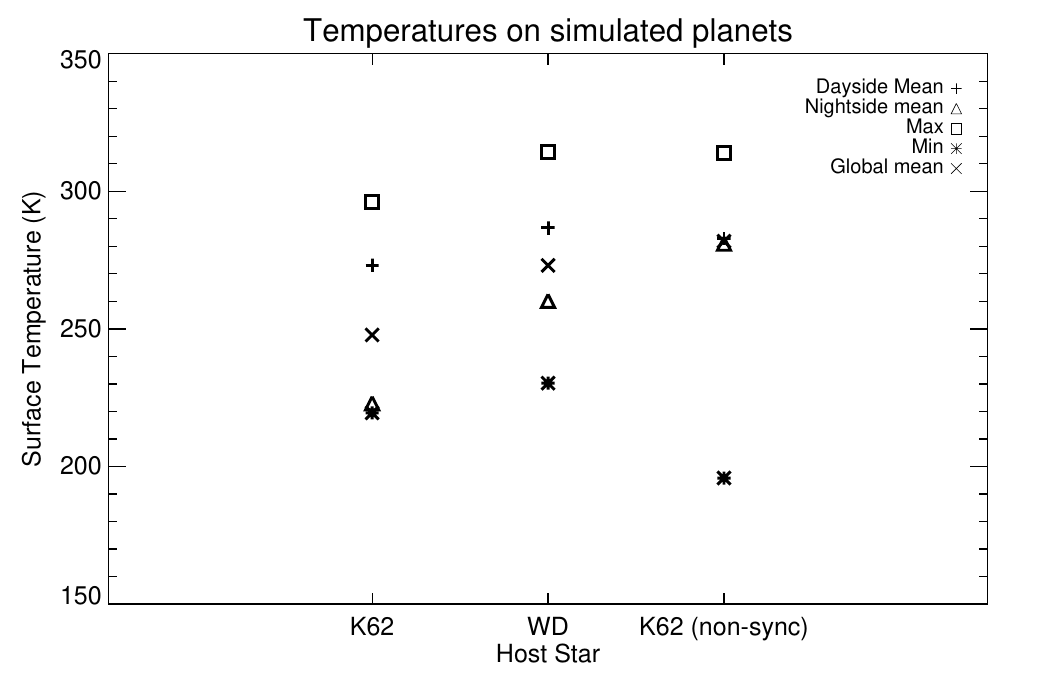}
\caption{While the non-synchronous planet exhibits a larger difference between maximum and minimum surface temperatures compared to either of the synchronous planets, with colder upper-latitude regions given the amount of ice exposed to sunlight throughout the planet's day, it also has global mean, dayside mean, and nightside mean surface temperatures that are nearly equivalent to each other, in large part due to a lower relative amount of reflective cloud cover and a strong nightside greenhouse effect.} 
 \label{Figure 9.}
\end{center}
\end{figure}

\section{Discussion} \label{sec:disc}
Our comparison of the climates of planets with Earth-like atmospheres and instellations orbiting a main-sequence star and a white dwarf with a similar effective temperature shows remarkable differences resulting from the environments hosting these planets. While minor variations are caused by the small difference in the effective temperatures of the two host stars, which affects the albedos of surface ice and snow on the planets, the major differences in resulting climates are due to the vastly different rotation periods of the planets, which affect their planetary large- and small-scale atmospheric circulation, cloud concentration and distribution, and strength of the greenhouse effect.

The WD planet is much warmer than the synchronous K62 planet, with a global mean surface temperature that is $\sim$25 K higher despite having the same solar constant, similar host star SED, and a similar cloud fraction. This contrast is due in part to the increased albedo of thicker, liquid water clouds present on the dayside of the K62 planet, demonstrating the impact of the dayside cloud contribution, given that liquid water clouds contribute most to a planet's TOA albedo \citep{Stephens2015}. This higher TOA albedo induces the reflection of more SW back to space, cooling temperatures, as indicated by the increased SWCF relative to that of the WD planet. Additionally, the much less efficient cloud and water vapor greenhouse effect on the nightside of the synchronous K62 planet, which has starker, cooler temperatures and a sharply lower mass of ice water clouds, which can contribute more strongly to a planet's greenhouse effect \citep{Mitchell1989, Borduas2018, Bjordal2020}, allows more radiation to escape to space, cooling the planet further. 

The surface temperature pattern seen for the WD planet in our study can be called, as has been termed in previous work, a ``bat rotator" pattern, constituting a regime of ultra-fast orbital period ($P <$ 1 day) within which most WD HZ planets are likely to occupy around their stars \citep{Zhan2024}. This much faster relative rotation and orbital period for the synchronous WD planet stretches out the clouds in the atmosphere, preventing thicker clouds from forming in the same manner as on the K62 planet, generating warmer dayside surface temperatures. K62's slower rotation creates much slower zonal winds and wind convergence at the substellar point, which is the traditional pattern expected on slow, synchronously-rotating planets. This pattern allows more clouds to form at the substellar point (see, e.g., \citealp{Way2018, Guzewich2020}), reflecting more radiation away from the planet via a strong relative stabilizing cloud feedback, often discussed as potentially beneficial for climate, by buffering close-in planets against runaway greenhouse states \citep{Yang2013, Yang2014, Way2018}. While a stabilizing cloud feedback may have a positive effect on planets near the inner edge of their host stars' habitable zones, where they receive high amounts of instellation that could subject their planets to such runaway greenhouse states, its potential advantage for planets that orbit squarely in the middle of their stars' habitable zones is lessened, as the net result of this cooling may be to reduce habitable surface area on the planet compared to planets with weaker (smaller net negative) SWCF. Planets occupying an intermediary, ``Rhines" rotational regime (5 days $< P <$ 20 days, \citealp{Haqq-Misra2018}) between the WD K62 planet's ultra-fast rotation and the K62 planet's slow rotation would likely have a stronger stabilizing cloud feedback than that of WD planets given a sharper day-night contrast in atmospheric circulation and resulting larger amount of substellar cloud cover, and occupy a middle-range in terms of the advantages and disadvantages of the stabilizing cloud feedback mechanism.

The difference in the amount of surface heating between the two planets certainly impacts habitable surface area, as shown by the ice fraction comparison on both planets (see Table 2). The increased relative heating and lower resultant ice fraction on the WD planet lead to a more optimistic likelihood of deglaciating any frozen planet that may have migrated into the HZ from farther out after the red giant phase (see, e.g., \citealp{Debes2002}) compared to a frozen synchronously-rotating planet orbiting a main-sequence star at an equivalent stellar flux distance, particularly if the stabilizing cloud feedback on such slower-rotating synchronous planets, which provides further cooling at close orbital distances \citep{Yang2013}, is taken into account. 

Surface temperatures reach far below the freezing point of liquid water on the K62 planet's nightside, likely resulting in the condensation of atmospheric water vapor onto the surface. Long-term evolution of synchronous rotators whose surfaces reach these temperatures could involve the eventual sequestration of water vapor content as ice that could form large glaciers \citep{Turbet2016}. However, depending on the planet's gravity and specific properties of the ice \citep{Leconte2013b}, as well as the planet's geothermal heat flux, which could cause basal melting \citep{Menou2013}, such glaciers could migrate to warmer regions of the planet where melting and sublimation back into the atmosphere could occur.

While a warmer surface environment could be more beneficial for life on a WD planet, the closer orbital distances of WD HZ planets, and relatively weaker stabilizing cloud feedback due to their ultra-fast rotation, may result in a greater susceptibility to a runaway greenhouse state and associated loss of surface water inventory characteristic of the inner Solar System planet Venus \citep{Ingersoll1969}. The magnitude of potential water loss on WD planets would ultimately depend on initial water inventory and the particular WD's luminosity evolution, as well as the degree of tidal heating (see, e.g., \citealp{Barnes2013a}). However, while their shorter orbital and rotation periods reduce dayside cloud coverage and planetary albedo relative to synchronous planets orbiting main-sequence hosts \citep{Yang2014, Kopparapu2016}, positioning the runaway greenhouse limit farther away from the star than for slower-rotating synchronous planets orbiting main-sequence stars, WD HZ planets would still rotate synchronously in all likelihood, thereby possessing some degree of stabilizing cloud feedback that reflects incoming SW radiation away from the planet. This latter, shared characteristic between WD HZ and main-sequence synchronous planets (though operating to a weaker degree on WD planets) would place the WD runaway greenhouse limit closer in to the star than for main-sequence stars hosting non-synchronous, rapidly-rotating planets, resulting in a wider HZ for WDs than originally surmised from 1-D studies \citep{Zhan2024}. Additionally, moist greenhouse states are less likely on WD planets with ultra-fast rotation periods as that in our study, due to drier upper atmospheres compared to planets orbiting main-sequence stars (see, e.g. \citealp{Zhan2024}), as suggested by the lower amount of relative cloud cover shown in Figure 4A. The susceptibility of WD planets to these aforementioned extremes of climate state at close orbital distances would therefore seem to exist within a similar regime space to that of planets orbiting main-sequence hosts given the somewhat balancing effects of spin synchronization and rotation period. 

We assumed a fixed, Earth-like atmosphere in our simulations. Given the early luminous phases of WDs, photolysis of surface H$_2$O may occur as mentioned above, resulting in the lighter hydrogen escaping to space and the heavier oxygen remaining behind in the atmosphere, as has been proposed to occur on M-dwarf planets during the pre-main-sequence phases of their host stars (see, e.g., \citealp{Luger2015b}). Such oxygenated atmospheres, if present on WD planets, could give rise to increased O$_3$ production (see, e.g., \citealp{Cooke2023}). Previous work exploring the potential impact of the UV environment of WDs on simulated orbiting habitable-zone planets found a substantial decrease in ozone column depth on a hypothetical planet orbiting a 5000 K WD at an equivalent flux distance to that of the Earth around the Sun, resulting in increased UVC radiation reaching the planet's surface, which could be harmful for life \citep{Kozakis2018}. However, that study was performed using a 1D radiative-convective model, and did not incorporate what is likely to be a synchronous rotational state for a planet orbiting in the habitable-zone of a WD star, thus garnering much higher global mean surface temperatures for their simulated planet than we find in our study of a synchronous WD HZ planet. Other work exploring the effects of different O$_2$ concentrations on atmospheric chemistry\textemdash finding reductions in CO$_2$, N$_2$O, CH$_4$, O$_3$, and water vapor column, and subsequent cooling of the troposphere when O$_2$ levels are reduced \citep{Cooke2023}\textemdash assumed 24-hr rotation periods. As ozone reactions are temperature sensitive (see, e.g., \citealp{Coates2016}), it would be interesting to see how an accurate surface and atmospheric temperature profile of a WD HZ planet would influence its ozone column depth and shielding for potential life, as greater ozone production would be expected at higher temperatures \citep{Coates2016}. Additionally, if greenhouse gases such as CO$_2$ and CH$_4$ are present in higher concentrations on WD HZ planets given their potentially oxygenated atmospheres, the additional atmospheric absorption is likely to result in higher surface temperatures than we calculated here for all planets. We would expect the atmospheric heating to be slightly higher on the K62 planet, due to its host star's slightly cooler effective temperature, resulting in a small increase in the near-IR radiation contribution to its spectrum relative to the WD (see Fig. 1) and associated atmospheric greenhouse gas absorption. Increased greenhouse gas concentrations could further buffer WD planets against global glaciations, though may be more advantageous for non-synchronous planets orbiting main-sequence stars farther out in their host stars' habitable zones. 

For main-sequence stars, the range of climates possible for habitable-zone planets is far greater given the many potential rotational-orbital spin states, as shown by our comparison of the synchronous and non-synchronous planet simulations. What seems clear is that a habitable-zone planet found around a WD star is more likely to be warm given its expected synchronous spin state. A non-synchronous habitable-zone planet orbiting a main-sequence star, despite an equally fast rotation period, is likely to carry some climatic advantages if its orbital period is much longer, including a lower amount of reflective cloud cover, a stronger nightside greenhouse effect, and lower-albedo ice surfaces if its host star has an even slightly cooler effective temperature, all of which contribute to a warmer environment than that of the (synchronously-rotating) WD HZ planet, though with colder poles and upper-latitude regions given the amount of ice exposed to sunlight throughout the planet's day in the non-synchronous case. However, the WD planet's fast 1:1 spin-orbit period, which generates strong and phase tilted meridional flux of zonal momentum and extended scales of cloud cover and atmospheric circulation, would seem to narrow this climatic gap compared to that between non-synchronous and synchronous planets orbiting a main-sequence star (see Table 2).

As it is likely that many of the planets orbiting WD progenitors will have been engulfed during the red giant phase, WD planets may be few within their systems, and possibly orbiting alone in single-planet systems. Other work has shown that (previously) spin-synchronized planets orbiting at the outer edge of the HZ in compact multiple-planet systems are more susceptible to global-scale glaciations, due to gravitationally-induced libration of the substellar point away from open ocean basins on these planets \citep{Chen2023}. Chen \textit{et. al} (\citeyear{Chen2023}) found that planets orbiting closer in to their stars within the HZ of these compact systems are less susceptible to these substellar longitude migrations resulting from gravitational perturbations by planetary companions, and that these effects are, naturally, nonexistent in single-planet systems. As other recent work has found planets in the 0.1-2$R_\Earth$ range orbiting in the HZ of WDs to be an order of magnitude lower in occurrence than larger (2-20$R_\Earth$) planets \citep{Kipping2024}, and evidence of circumstellar disk material around WDs could be indicative of the tidal disruption of any planets that did originally survive the red giant phase (see, e.g., \citealp{Aungwerojwit2024}), WD HZ planets may indeed be few or single around WDs, and therefore more robust against such snowball states, due to their extremely close orbital distances given their host stars' luminosities, which all but ensure a consistently fixed substellar point.
 
While the 10-hr synchronous rotational and orbital period of our hypothetical WD planet lies at the other extreme compared to that of a synchronous habitable-zone planet orbiting a main-sequence star, the stretched out scales often seen in this fast-rotating regime in simulations of terrestrial planets (see, e.g., \citealp{Showman2011, Kaspi2015, Haqq-Misra2018, Komacek2019a, Guzewich2020} ), and comparatively diminished SWCF and enhanced LWCF pose the advantage of allowing such planets to retain more of the heat from their stars, even as their expired nuclear furnaces continue to cool over time. However, an additional counterpoint to this potential advantage exists, as early in the evolutionary stage of a white dwarf, when it is more luminous, orbiting close-in planets that are already warm would be rendered even hotter, which could reduce habitable surface area. Our understanding of the ultimate habitability potential of WD planets would benefit from future studies that incorporate the time-evolving habitable zone of a white dwarf to explore the evolution of a rocky planet's climate sensitivity as its host star's stellar flux decreases over time, though this would require the incorporation of a carbonate-silicate cycle, which regulates silicate weathering and planetary surface temperature \citep{Walker1981}. Regardless, the results of the study presented here suggest that the white dwarf stellar environment, once dismissed as unimaginable for life, may present a newfound consideration for observational efforts in search of habitable planets in the coming decades. 

\section{Conclusions} \label{sec:conc}
We used a 3D GCM to simulate the climates of synchronously-rotating planets with Earth-like atmospheres and instellations orbiting in the habitable zones of a 5000 K white dwarf star and main sequence K-dwarf star Kepler-62 of similar effective temperature. We have shown their markedly differing climates to be largely due to the rotation periods of the two planets at their respective orbital distances. While the much slower, 155-day rotation and orbital period of the planet orbiting Kepler-62 exhibits a large dayside liquid water cloud mass, shortwave cloud forcing, and top-of-atmosphere albedo, along with a weakened nightside greenhouse effect, the WD planet's much faster, 10-hr rotation and orbital period generates strong zonal winds that stretch out the atmospheric and cloud circulation across the planet, fewer dayside liquid water clouds, and a stronger greenhouse effect on the nightside, contributing to a global mean surface temperature that is $\sim$25 K higher than that of the K62 planet. In the non-synchronous case for the planet orbiting Kepler-62 with the 10-hr rotation period of the WD planet, the larger fraction of surface ice that is exposed to sunlight over all longitudes, though lower-albedo ice compared to that of the WD planet, generates an overall warmer surface environment, albeit with lower temperatures in the upper latitudes and at the poles. That the strong and homogeneous atmospheric circulation of the WD planet compensates for much of the increased warming on the non-synchronous planet, has important implications for future exoplanet characterization efforts. The potential climates of planets that formed within or migrated to the habitable zones of white dwarf stars may be more accurately assessed prior to direct observational confirmation of their spin states, as they are more likely to exist in a synchronous rotation state, provide warmer surface temperatures than synchronous planets orbiting main-sequence stars, and comparable surface temperatures even to non-synchronous, rapidly-rotating planets orbiting main-sequence stars. WD habitable-zone planets may therefore harbor more clement conditions for life to compensate for the cooling and dimming of their host stars over time. 

\section{Acknowledgments} \label{sec:acknowl}

This material is based upon work supported by NSF grant No. 1753373, and by a Clare Boothe Luce Professorship. We would like to acknowledge high-performance computing support from Cheyenne (doi:10.5065/D6RX99HX) and Derecho (doi:10.5065/qx9a-pg09) provided by the National Science Foundation (NSF) National Center for Atmospheric Research. A.S. thanks Yusra AlSayyad for coding help with white dwarf grid spectral models, Stephen Warren for helpful communications regarding the emissivities of icy surfaces, and the U See I Write retreat program at UC Irvine for support in the preparation of this manuscript. ETW acknowledges NASA Habitable Worlds Grant No. 80NSSC20K1421. EA acknowledges support from NSF grant No. AST-1907342, NASA NExSS grant No. 80NSSC18K0829, and NASA XRP grant No. 80NSSC21K1111. P.-E. T. acknowledges support from the European Research Council (ERC) under the European Union's Horizon 2020 research and innovation programme (grant agreement no. 101002408).





\end{document}